\documentclass{emulateapj}
\shorttitle{A young massive star cluster in M31} \shortauthors{Ma et al.}

\begin{document}
\slugcomment{AJ, in press}
\title{Age and mass constraints for a young massive
cluster in M31 based on spectral-energy-distribution fitting}

\author{Jun Ma,\altaffilmark{1,2} Song Wang,\altaffilmark{1,3} Zhenyu Wu,\altaffilmark{1} Zhou
  Fan,\altaffilmark{1} Yanbin Yang,\altaffilmark{1} Tianmeng Zhang,\altaffilmark{1} Jianghua
  Wu,\altaffilmark{1} Xu Zhou,\altaffilmark{1} Zhaoji Jiang,\altaffilmark{1} and Jiansheng
  Chen\altaffilmark{1}}

\altaffiltext{1}{National Astronomical Observatories, Chinese Academy
  of Sciences, Beijing 100012, P. R. China;\\ majun@nao.cas.cn}

\altaffiltext{2}{Key Laboratory of Optical Astronomy, National
  Astronomical Observatories, Chinese Academy of Sciences, Beijing
  100012, P. R. China}

\altaffiltext{3}{Graduate University, Chinese Academy of Sciences,
  Beijing 100039, P. R. China}

\begin{abstract}
VDB0-B195D is a massive, blue star cluster in M31. It was observed as
part of the Beijing-Arizona-Taiwan-Connecticut (BATC) Multicolor Sky
Survey using 15 intermediate-band filters covering a wavelength range
of 3000--10,000 \AA. Based on aperture photometry, we obtain its
spectral-energy distribution (SED) as defined by the 15 BATC filters.
We apply previously established relations between the BATC
intermediate-band and the Johnson-Cousins $UBVRI$ broad-band systems
to convert our BATC photometry to the standard system.  A detailed
comparison shows that our newly derived $VRI$ magnitudes are fully
consistent with previous results, while our new $B$ magnitude agrees
to within $2\sigma$. In addition, we determine the cluster's age and
mass by comparing its SED (from 3000 to 20,000{\AA}, comprising
photometric data in the 15 BATC intermediate bands, optical broad-band
$BVRI$, and 2MASS near-infrared $JHK_{\rm s}$ data) with theoretical
stellar population synthesis models, resulting in age and mass
determinations of $60.0\pm 8.0$~Myr and $(1.1-1.6) \times 10^5
M_\odot$, respectively. This age and mass confirms previous
suggestions that VDB0-B195D is a young massive cluster in M31.
\end{abstract}


\section{Introduction}

Young massive star clusters (YMCs) are among the main objects
resulting from violent star-forming episodes triggered by galaxy
collisions, mergers, and close encounters \citep[see][and references
therein]{grijs07}. They are also referred to as `young populous
clusters,' a term first coined by \citet{hodge61}, who used it to
describe 23 clusters containing bright, blue stars in the Large
Magellanic Cloud. In \citet{hodge61}, the `young' aspect is
demonstrated by the fact that all clusters have main sequences that
extend to absolute magnitudes brighter than $M_V=0$, while `populous'
describes their richness (stellar membership). However, YMCs are also
observed in quiescent galaxies \citep{lr99} and in the disks of
isolated spirals, although higher cluster-formation efficiencies are
associated with environments exhibiting high star-formation rates
\citep[see][and references therein]{larsen04a,cw07}. It has become
clear that, in many ways, YMCs resemble young versions of the old
globular clusters (GCs) associated with all large galaxies
\citep[see][and references therein]{larsen04b}. YMCs are seemingly
absent in the Milky Way; possibly the best example of a Galactic YMC
is Westerlund 1, a heavily reddened cluster with an age and mass of
4--5 Myr \citep{Crowther06} and $M_{\rm cl} \sim 10^5~M_\odot$
\citep{Clark05}, respectively.

Since the pioneering work of \citet{Tinsley68,Tinsley72} and
\citet{SSB73}, evolutionary population synthesis modeling has become a
powerful tool to interpret integrated spectrophotometric observations
of galaxies and their components, such as star clusters
\citep[e.g.,][]{Anders04}. The evolution of star clusters is usually
modeled by means of the simple stellar population (SSP)
approximation. An SSP is defined as a single generation of coeval
stars formed from the same progenitor molecular cloud (thus implying a
single metallicity), and governed by a given stellar initial mass
function (IMF).

Age and metallicity are two basic star cluster parameters. The most direct method to determine a
cluster's age is by employing main-sequence photometry, since the absolute magnitude of the
main-sequence turnoff is predominantly affected by age \citep[see][and references
therein]{puzia02}. However, until recently \citep[cf.][]{perina09}, this method was only applied
to the star clusters in the Milky Way and its satellites \citep[e.g.,][]{rich01}, although
\citet{brown04} estimated the age of an M31 GC using extremely deep images observed with the {\sl
Hubble Space Telescope (HST)}'s Advanced Camera for Surveys. Generally, the ages of extragalactic
star clusters are determined by comparing their observed spectral-energy distributions (SEDs)
and/or spectroscopy with the predictions of SSP models
\citep{Williams01a,Williams01b,degrijs03a,degrijs03b,degrijs03c,bik03,jiang03,Beasley04,puzia05,ma06,fan06,ma07a,ma09,cald09,
Wang10}. Nevertheless, SSP models assume that cluster IMFs are fully populated, i.e., that
clusters contain infinite numbers of stars with a continuous distribution of stellar masses, and
that all evolutionary stages are well sampled. Real clusters, however, contain a finite number of
stars. Therefore, a disagreement between the observed cluster colors and theoretical colors
derived from SSP models may become apparent \citep[see][and references therein]{pisk09,ph10}.
Other limitations inherent to SSP models arise from our poor understanding of some advanced
stellar evolutionary stages, such as the supergiant and the asymptotic-giant-branch (AGB) phases
\citep[see][and references therein]{bc03}.

Located at a distance of $785 \pm 25$ kpc, corresponding to a distance
modulus of $(m-M)_0 = 24.47 \pm 0.07$ mag \citep{McConnachie05}, M31
is the nearest and largest spiral galaxy in the Local Group of
galaxies. It has been the subject of many GC studies and surveys,
dating back to the early study of \citet{hubble32}. Based on previous
publications \citep{hubble32,seynas45,hilt58,mayegg53,km60},
\citet{vete62a} compiled the first large M31 GC catalog, containing
$UBV$ photometric data of approximately 300 GC candidates. Over the
past decades, several major catalogs of M31 GCs and GC candidates have
been published, including major efforts by the Bologna group
\citep{batt80,batt87,batt93}, \citet{bh00},
\citet{gall04,gall05,gall06,gall07}, \citet{kim07}, \citet{cald09},
and \citet{peacock09}.  Following on from the first extensive
spectroscopic survey of M31 GCs by \citet{Sidney69}, a significant
number of authors \citep[e.g.,][and references
therein]{hsv82,hbk91,DG,Federici93,jab,bh00,per02,gall06,lee08} have
studied their spatial, kinematic, and chemical (metallicity)
properties.

M31 is known to host a large number of young star clusters
\citep[e.g.,] [and references
therein]{Fusi05,cald09,Wang10}. \citet{Fusi05} presented a
comprehensive study of 67 very blue star clusters, which they referred
to as `blue luminous compact clusters' (BLCCs). Since they are quite
bright ($-6.5\leq M_V \leq -10.0$ mag) and very young ($<2$ Gyr),
BLCCs may be equivalent to YMCs \citep[see for
details][]{perina09,perina09b}. To ascertain their properties,
\citet{perina09,perina09b} performed an imaging survey of 20 BLCCs in
the disk of M31 using the {\sl HST}'s Wide Field and Planetary
Camera-2 (WFPC2). They obtained the reddening values, ages, and
metallicities of their sample clusters by comparing the observed
color-magnitude diagrams (CMDs) and luminosity functions with
theoretical models.

VDB0-B195D was first detected by \citet{Sidney69}. Its color is
extremely blue \citep[e.g., $U-B=-0.48$ mag;][]{Sidney69} and it is
very bright in blue bands \citep[e.g., $U=14.66$ mag;][]{Sidney69}. As
a consequence, \citet{Sidney69} asserted that VDB0-B195D is the
brightest open cluster in M31. He determined an integrated stellar
spectral type equivalent to A0, which implies that the cluster
contains massive stars. In addition, VDB0-B195D is particularly
extended and most previous photometric studies did not include the
full extent of the object's light distribution \citep[see for
details][]{perina09}. We will provide an overview of previous studies
that included the cluster in \S 2.1. It was observed as part of the
galaxy calibration program of the Beijing-Arizona-Taiwan-Connecticut
(BATC) Multicolor Sky Survey \citep[e.g.,][]{fan96,zheng99} in 15
intermediate-band filters. Combined with photometry in optical
broad-band $BVRI$ and near-infrared $JHK_{\rm s}$ filters from the Two
Micron All Sky Survey (2MASS) taken from \citet{perina09}, we
obtained the SED of VDB0-B195D in 22 filters, covering the wavelengh
range from 3000 to 20,000 \AA.

In this paper, we describe the details of the observations and our
approach to the data reduction in \S 2. In \S 3, we determine the age
and mass of VDB0-B195D by comparing observational SEDs with population
synthesis models. We discuss the implications of our results and
provide a summary in \S 4.

\section{Optical and near-infrared observations of the YMC VDB0-B195D}

\subsection{Historical overview}

VDB0-B195D was first given the designation `0' (i.e., VDB0), the
brightest open cluster in M31, in the catalog of
\citet{Sidney69}. \citet{batt87} identified VDB0-B195D independently
and called it B195D. In \citet{batt87}, B195D was given a low level of
confidence (class D) of being a genuine cluster (classes A and B were
assigned very high and high levels of confidence, respectively).  It
was only recently independently confirmed to be a single
object. \citet{cald09} presented a new catalog containing 670 likely
star clusters, stars, possible stars, and galaxies in the field of
M31, all with updated high-quality coordinates accurate to $0.2''$,
based on images from either the Local Group Galaxies Survey (LGGS)
\citep{massey06} or the Digitized Sky Survey (DSS). They use the
designation VDB0-B195D, associated with $\rm \alpha_0=00^{\rm
h}40^{\rm m}29^{\rm s}.43$ and $\rm \delta_0=+40^{\circ}36'14''.8$
(J2000.0), which are the coordinates we adopt in this
paper. Independently, \citet{perina09} studied the properties of
VDB0-B195D in detail based on their {\sl HST}/WFPC2 imaging survey of
young massive GCs in M31. They initially selected VDB0-B195D as two
YMCs in M31, but their WFPC2 images showed unequivocally that these
two sample objects are, in fact, the same cluster. In addition, the
{\sl HST} images clearly confirmed that VDB0-B195D is a real
cluster. However, it is difficult to establish whether it is more
similar to ordinary open clusters, similar to those in the disk of the
Milky Way, than to YMCs that may evolve to become disk GCs \citep[see
for details][]{perina09}. Spectral observations of VDB0-B195D were
obtained by \citet{Sidney69}---yielding classification spectra and the
object's radial velocity---and \citet{per02}, who used them for
determination of its radial velocity and metallicity.

\subsection{Archival images of the BATC Multicolor Sky Survey}

Observations of the YMC VDB0-B195D were obtained with the BATC 60/90cm
Schmidt telescope located at the XingLong station of the National
Astronomical Observatory of China (NAOC). This telescope is equipped
with 15 intermediate-band filters covering the optical wavelength
range from 3000 to 10,000 \AA. The filter system was specifically
designed to avoid contamination by the brightest and most variable
night-sky emission lines. Descriptions of the BATC photometric system
can be found in \citet{fan96}. Before February 2006, a Ford Aerospace
2k$\times$2k thick CCD camera was installed, with a pixel size of 15
$\mu$m and a field of view of $58^{\prime}\times 58^{\prime}$,
yielding a resolution of $1.7''$ pixel$^{-1}$. Since February 2006, a
new E2V 4k$\times$4k thinned CCD with a pixel size of 12 $\mu$m has
been in operation, featuring a resolution of $1.3''$ pixel$^{-1}$. The
blue quantum efficiency of the new, thinned CCD is 92.2\% at 4000~\AA,
which is much higher than for the old, thick device \citep[see for
details][]{fan09}. A field including VDB0-B195D in the $a$--$c$
filters was observed with the thinned CCD, and in $d$--$p$ bands with
the thick CCD. Fig. 1 shows a finding chart of VDB0-B195D in the BATC
$g$ band (centered at 5795 \AA), obtained with the NAOC 60/90cm
Schmidt telescope. We adopt an aperture with a radius of $15''$ (shown
in Fig. 1) for the integrated photometry discussed in this paper.

\begin{figure*}
\figurenum{1} \hspace{1.0cm}{\plotone{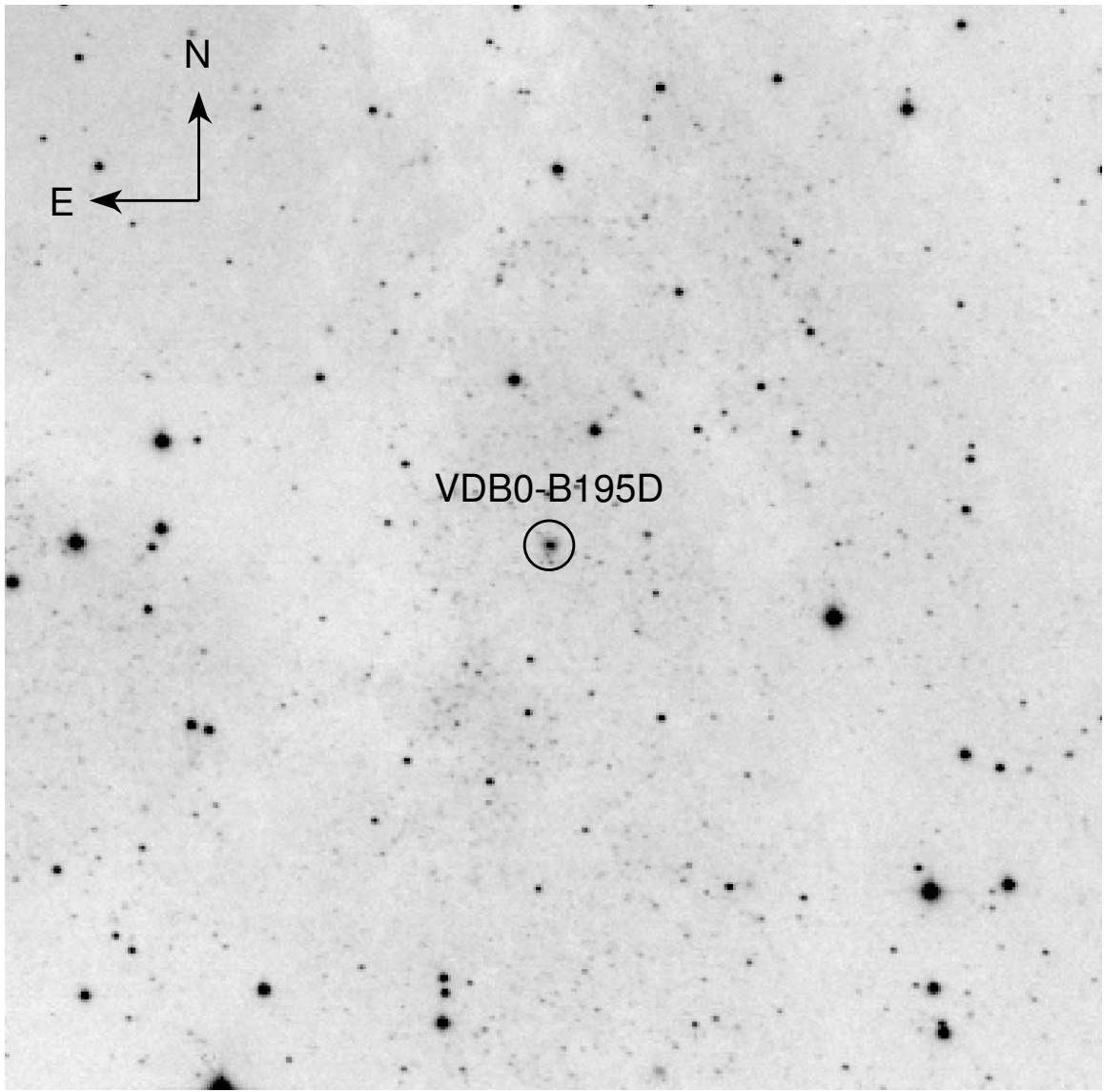}} \vspace{0.5cm}
\caption{Image of VDB0-B195D in the BATC $g$ band, obtained with the
NAOC 60/90cm Schmidt telescope. VDB0-B195D is circled using an
aperture with a radius of $15''$. The field of view of the image is
$11^{\prime}\times 11^{\prime}$.} \label{fig:one}
\end{figure*}

The BATC survey team obtained 61 images of VDB0-B195D in 15 BATC
filters between January 2004 and November 2006. \citet{fan09}
performed the data reduction of these images, which formed part of
their M31-7 field. Table 1 contains the observation log, including the
BATC filter names, the central wavelength and bandwidth of each
filter, the number of images observed through each filter, and the
total observing time per filter. Multiple images through the same
filter were combined to improve image quality (i.e., increase the
signal-to-noise ratio and remove spurious signal).

\subsection{Intermediate-band photometry of VDB0-B195D}

We determined the intermediate-band magnitudes of VDB0-B195D on the
combined images using a standard aperture photometry approach, i.e.,
the {\sc phot} routine in {\sc daophot} \citep{stet87}. Calibration of
the magnitude zero level in the BATC photometric system is similar to
that of the spectrophotometric AB magnitude system. For flux
calibration, the Oke-Gunn \citep{ok83} primary flux standard stars HD
19445, HD 84937, BD +26$^{\circ}$2606, and BD +17$^{\circ}$4708 were
observed during photometric nights \citep{yan00}. VDB0-B195D is
located in the M31-7 field of \citet{fan09}. The absolute flux of the
M31-7 field was calibrated based on secondary standard transformations
using the M31-1 field, which was calibrated, in turn, by the four
Oke-Gunn primary flux standard stars by \citet{jiang03}. Since
VDB0-B195D is an extended object, an appropriate aperture size must be
adopted to determine its total luminosity. The (radial) photometric
asymptotic growth curve, in all BATC bands, flattens out at a radius
of $\sim 15''$. Inspection ensured that this aperture is adequate for
photometry, i.e., VDB0-B195D does not show any obvious signal beyond
this radius. In addition, this aperture is nearly the same as that
adopted by \citet{perina09} to determine the cluster's photometry in
the $BVRI$ bands, based on the M31 imaging survey of \citet{massey06}
(see \S 2.4 below). Therefore, we use an aperture with $r\approx15''$
for integrated photometry, i.e., $r=9$~pixels for the 2k$\times$2k
thick CCD camera, and $r=12$~pixels for the 4k$\times$4k thinned CCD
camera. VDB0-B195D is projected onto the disk of M31, where the
background is bright and fluctuates, potentially as a function of
distance from the cluster center. To avoid contamination from
background fluctuations, we adopted annuli for background subtraction
spanning between 10 and 15 pixels for the 2k$\times$2k thick CCD
camera, and from 13 to 20 pixels for the 4$\times$4k thinned CCD
camera, both corresponding to $\sim 17$--26$''$. While these annuli
are spatially as close as possible to the region dominated by cluster
light (so that any differences in background flux are minimized), they
are wide enough to average out any expected background
fluctuations. The calibrated photometry of VDB0-B195D in 15 filters is
summarized in column (6) of Table 1, in conjunction with the
$1\sigma$ magnitude uncertainties, which include uncertainties
from the calibration errors of both the M31-1 field standard stars
\citep[see for details][]{fan09,jiang03} and `the secondary standard
stars' in common between the M31-1 and M31-7 fields used for
calculation of the mean magnitude offsets between the standard and
instrumental magnitudes \citep[see for details][]{fan09}, as well as
those resulting from our {\sc daophot} application.

\subsection{Optical broad-band and near-infrared 2MASS photometry of VDB0-B195D}

Four independent sets of photometric data exist for
VDB0-B195D. \citet{Sidney69} obtained $UBV$ photometry using
observations of the 200-inch Hale telescope, \citet{batt87} performed
$UBVR$ photometry based on photographic plates observed with the 152
cm Ritchey-Chr\'etien $f$/8 telescope of the University of Bologna in
Loiano, \citet{king91} obtained $UBV$ photometry for VDB0-B195D using
observations with the University of Hawaii's 2.2 m telescope on Mauna
Kea using the $f$/10 secondary and coronene-coated $584 \times 416$
GEC CCD, and \citet{sh95} performed $UBV$ photometry based on
photo-electric observations with the 2.6 m Shain telescope of the
Crimean Astrophysical Observatory. In addition, in the Revised Bologna
Catalogue (RBC) of M31 GCs published by \citet{gall04}, the
photometric data of VDB0-B195D in optical bands are based on
\citet{batt87} and \citet{sh95}, and transformed to the reference
system of \citet{bh00} by applying offsets derived from objects in
common between the relevant catalog and the data set of
\citet{bh00}. In the RBC, VDB0-B195D was regarded as two objects. We
list these photometric data in Table 2 for comparison. Note that, in
the latest RBC incarnation (version 3.5, updated on 27 March 2008),
VDB0-B195D is included as a single object.

\citet{gall04} also determined 2MASS $JHK_{\rm s}$ photometric
magnitudes for VDB0-B195D \citep[transformed to the CIT photometric
system;][]{Elias82, Elias83}, which we have included in Table 3. In
addition, \citet{perina09} realized that VDB0-B195D is a particularly
extended object and that it is possible that the photometry of
\citet{sh95} (compiled in the RBC) was obtained with apertures that
were not large enough to include all of its flux. Therefore, they
redetermined its photometric values in the $BVRI$ bands based on the
M31 imaging survey of \citet{massey06} using an aperture with
$r=14.4''$, which are also listed in Table 3.

From a comparison of the values in Tables 2 and 3, it is clear that
the magnitudes of \citet{Sidney69} are brighter, while the results of
the three other references are consistent. The magnitudes determined
by \citet{perina09} are much brighter, however, because of their
careful inclusion of all of the cluster's flux. To compare our
photometric results with previously published values, we transformed
the magnitudes of VDB0-B195D in the BATC intermediate bands to
broad-band $UBVRI$-equivalent photometry based on the relationships
obtained by \citet{zhou03}.  These are also listed in Table 3,
and the uncertainties include those originating from the
transformation based on the relationships of \citet{zhou03} and their
calibration errors (column 5 of their Table 3). In Fig. 2, we show
the result of the comparison. In general, the other photometric data
are fainter than ours and those of \citet{perina09}. Fig. 2 and Table
3 show that our new $VRI$ magnitudes agree with the results of
\citet{perina09}, and that the $B$ magnitude obtained in this paper is
0.32 mag brighter than that of \citet{perina09}. Considering the
photometric errors of both \citet{perina09} and our current study,
these two $B$-band photometric results are consistent within
$2\sigma$. In addition, we should keep in mind that, although the
$VRI$ magnitudes obtained in this paper are consistent with the
results of \citet{perina09} within $1\sigma$, the disagreement in $B$
magnitudes at this level is understandable. This is caused by the fact
that the original photometry in the present paper was obtained in the
proprietary BATC filters and transformed to the $UBVRI$ system using
transformation equations. \citet{zhou03} determined these
conversions based on the broad-band $UBVRI$ magnitudes of 48 stars
from \citet{lan83,lan92} and \citet{GE2000} in the Landolt SA95 field,
and their photometric data in the 15 BATC intermediate-band filters.
In addition, the central wavelengths and bandwidths of the BATC and
$UBVRI$ systems differ. In fact, a similar significant disagreement of
$B$-band photometric data for some M31 GCs was reported by
\citet{Wang10}, citing similar arguments.

\begin{figure*}
\figurenum{2} \hspace{1.0cm}\epsscale{0.75}\hspace{0.0cm}\rotatebox{-90}{\plotone{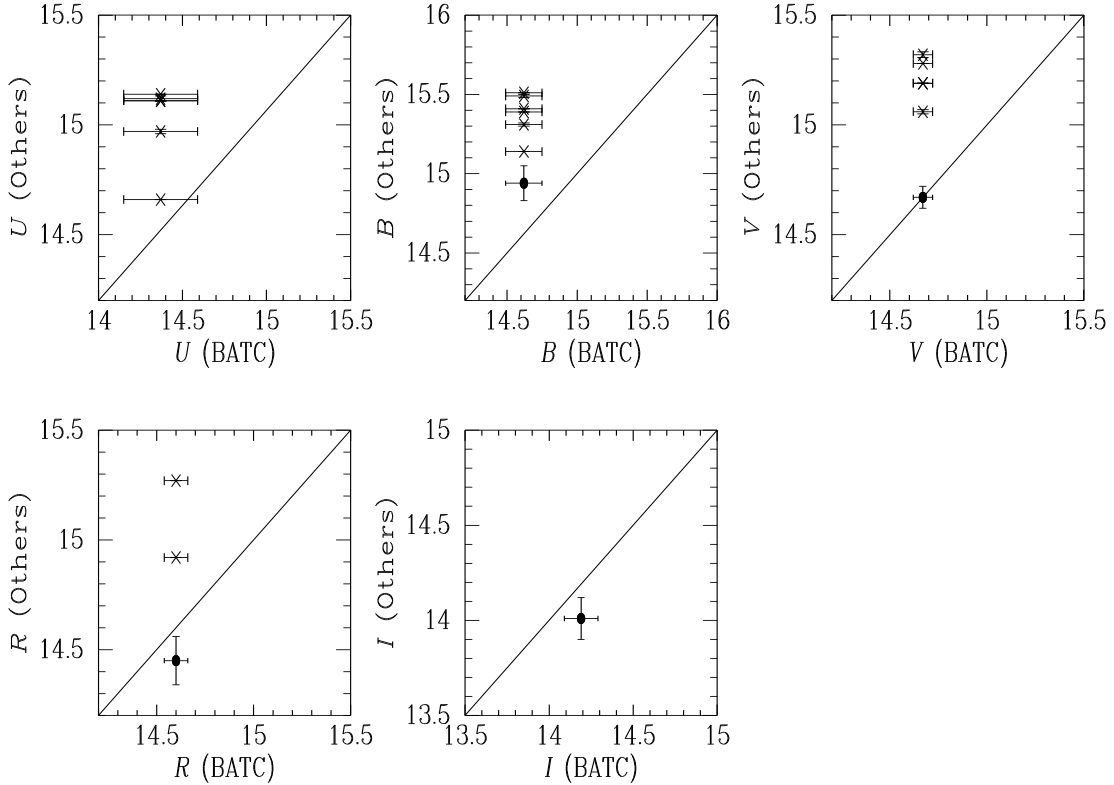}}
\vspace{-0.6cm} \caption{Comparison of photometric data from different
sources with new determinations in this paper for VDB0-B195D. The data
points shown as black dots are from \citet{perina09}.} \label{fig:two}
\end{figure*}

\section{Stellar population of VDB0-B195D}

\subsection{Stellar populations and synthetic photometry}

To determine the age and mass of VDB0-B195D, we compared its SED with
theoretical stellar population synthesis models. The SED consists
of photometric data in the 15 BATC intermediate bands obtained in this
paper and optical broad-band $BVRI$ and 2MASS near-infrared $JHK_s$
data from \citet{perina09}, listed in Table 3. We used the {\sc
galev} SSP models \citep[e.g.,][]{kurth99,schulz02,anders03} for our
comparisons. The {\sc galev} SSPs are based on the Padova stellar
isochrones, with the most recent versions using the updated
\citet{bertelli94} isochrones (which include the thermally pulsing
asymptotic giant-branch phase), and a \citet{salp55} stellar IMF with
lower- and upper-mass limits of 0.10 and between 50 and 70 $M_\odot$,
respectively, depending on metallicity. The full set of models spans
the wavelength range from 91{\AA} to 160 $\mu$m.  These models cover
ages from $4 \times 10^6$ to $1.6 \times 10^{10}$ yr, with an age
resolution of 4 Myr for ages up to 2.35 Gyr, and 20 Myr for greater
ages. The {\sc galev} SSP models include five initial metallicities,
$Z=0.0004, 0.004, 0.008, 0.02$ (solar metallicity), and 0.05.

Since our observational data consist of integrated luminosities
through the set of BATC filters, we convolved the {\sc galev} SSP SEDs
with the BATC intermediate-, optical broad-band $BVRI$, and 2MASS
filter-response curves to obtain synthetic optical and near-infrared
photometry for comparison. The synthetic $i^{\rm th}$ filter magnitude
can be computed as
\begin{equation}
m=-2.5\log\frac{\int_{\nu}F_{\nu}\varphi_{i} (\nu){\rm d}\nu}{\int_{\nu}\varphi_{i}(\nu){\rm
d}\nu}-48.60,
\end{equation}
where $F_{\nu}$ is the theoretical SED and $\varphi_{i}$ the response curve of the $i^{\rm th}$
filter of the BATC, $BVRI$, and 2MASS photometric systems. Here, $F_{\nu}$ varies with age and
metallicity. Since the observed magnitudes in the $BVRI$ and 2MASS photometric systems are given
in the Vega system, we transformed them to the AB system for our fits.
%The photometric offsets in
%the $BVRI$ and 2MASS filters between the Vega and AB systems were obtained based on equations (7)
%and (8) in the manual provided by \citet{bc03} (bc03.ps).

\subsection{Reddening and metallicity of VDB0-B195D}

To obtain the intrinsic SED of VDB0-B195D, its photometry must be
dereddened. To date, only \citet{perina09} obtained reddening values
for VDB0-B195D. They compared the observed CMD with theoretical
isochrones and determined $E(B-V)=0.20\pm0.03$ mag. \citet{cald09}
were unable to derive the cluster's reddening value because of the
presence of a foreground field star, so they adopted $E(B-V)=0.28\pm
0.17$ mag (external rms error), equivalent to the mean reddening of
the young clusters in M31. In this paper, we therefore adopt the
reddening value from \citet{perina09}.

In addition, cluster SEDs are affected by age and metallicity
effects. Therefore, we can only accurately constrain a cluster's age
if the metallicity is known. \citet{perina09} found that the CMD of
VDB0-B195D, based on their {\sl HST}/WFPC2 observations, is best
reproduced by the solar-metallicity models of \citet{Girardi02}. We
therefore adopt solar metallicity for VDB0-B195D.

\subsection{The `lowest-luminosity-limit' test}

The lowest-luminosity limit \citep[LLL;][]{CL04} implies that it is
meaningless to compare a cluster with population synthesis models to
obtain its age and mass if its integrated luminosity is lower than the
luminosity of the most luminous star included in the model for the
relevant age.  The LLL method states that clusters fainter than this
limit cannot be analyzed using standard procedures such as $\chi^2$
minimization of the observed values with respect to the mean SSP
models \citep[see also][]{Barker08}. Below the LLL, cluster ages and
masses cannot be obtained self-consistently. To take into account the
effects on the integrated luminosities of statistically sampling the
stellar IMF \citep[e.g.,][]{CLC00,CVGLMH02,CL04}, we used the
theoretical Padova isochrones at http://stev.oapd.inaf.it/cmd
(CMD2.2). This interactive Web interface provides isochrones for a
number of photometric systems, including optical broad-band, 2MASS,
and the BATC data used here. We obtained the solar-metallicity
($Z=0.019$) isochrones of \citet{Marigo08}, as recommended by CMD2.2,
based on the \citep{salp55} IMF so as to match the IMF selection for
our age and mass determinations of VDB0-B195D in \S 3.4 (see \S 3.1
for details).

Figure 3 shows the LLL values as a function of age for the different
filters used in this paper. These luminosities are obtained by
identifying the most luminous star on each isochrone for the relevant
passband. The gray area shows the cluster's absolute luminosity,
assuming a distance modulus of $(m-M)_{0}=24.47$ mag (785 kpc) for M31
\citep{McConnachie05}. The upper luminosity limit has been corrected
for extinction, based on a reddening value of $E(B-V)=0.20$ mag. The
interstellar extinction curve, $A_{\lambda}$, is taken from
\citet{car89}, $R_{V}=A_V/E(B-V)=3.1$.

We see that VDB0-B195D does not lie below the LLL in any of the
passbands used here. This means that, in general, VDB0-B195D can host
the most luminous star that would be present theoretically for the
given age of the cluster.

\begin{figure*}
\figurenum{3} \hspace{1.0cm}\epsscale{1.2}\hspace{-1.2cm}\rotatebox{0}{\plotone{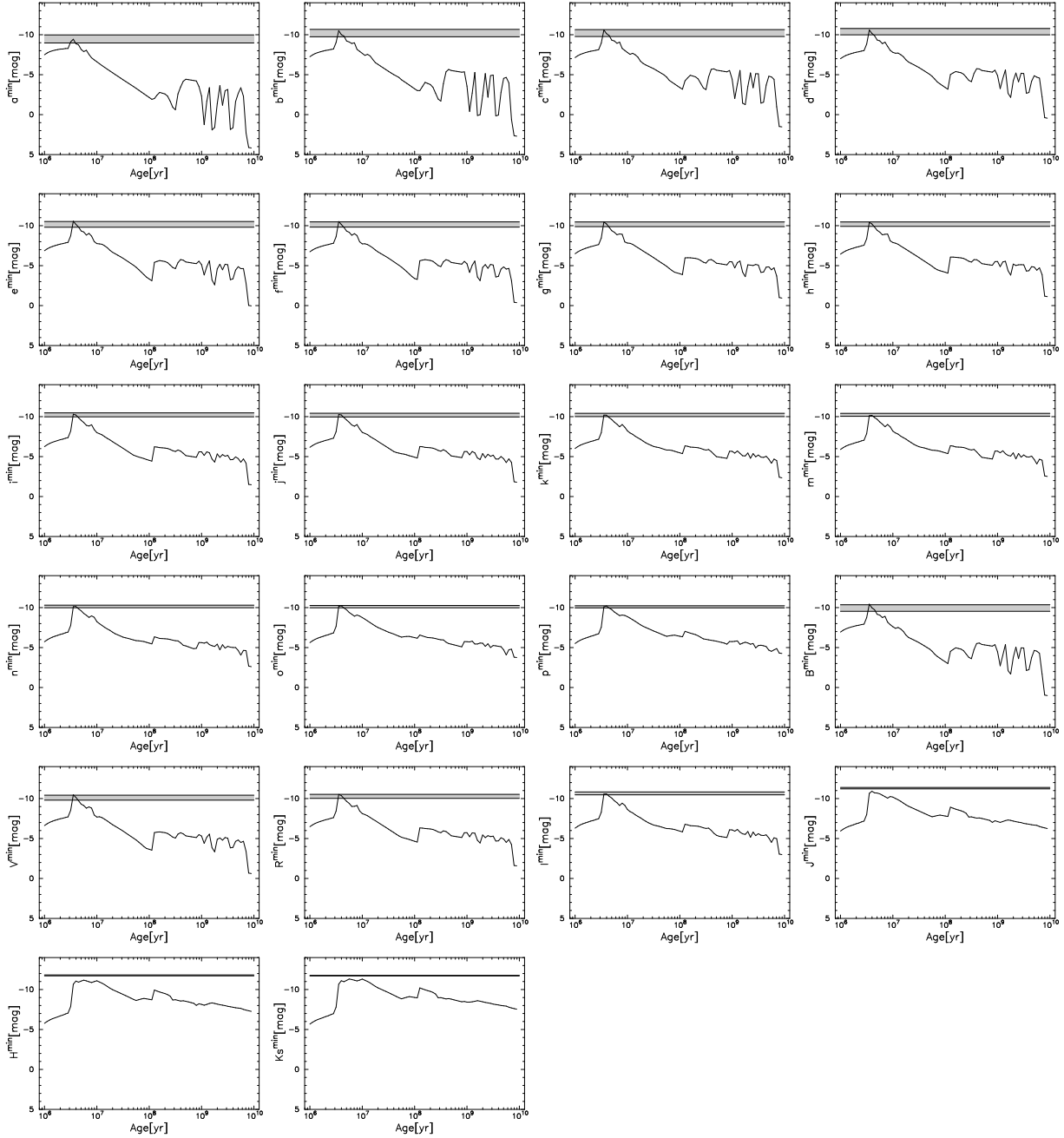}}
\vspace{-7.0cm} \caption{Lowest-luminosity limit for the filters used
in this paper. The curves indicate the luminosities of the most
luminous star on each isochrone for the relevant passband. The
light-gray area shows the absolute magnitudes of VDB0-B195D based on a
reddening value of $E(B-V)=0.20$ mag \citep{perina09}. We used a
distance modulus of $(m-M)_{0}=24.47$ mag (785 kpc) for M31
\citep{McConnachie05} to calculate the absolute magnitudes.}
\label{fig:three}
\end{figure*}

\subsection{Fit results}

In the previous section, the LLL test proves that the luminosity of VDB0-195D
is higher than the luminosity of its brightest star expected for a given cluster
age, i.e., that using SSP models is not completely meaningless. In addition, the bright absolute
magnitude of VDB0-195D allows us to consider a possibility that the cluster is massive
enough and IMF sampling effects should not strongly impact the fitting results. So we will
determine the cluster's age and mass estimates based on direct
comparisons with SSP mean values in this section. However, we should keep in mind that this
approach is a compromise. In fact, the fitting results (Fig. 4 and Table 5) show probable problem
even for relative massive clusters.

We use a $\chi^2$ minimization test to determine which {\sc galev} SSP
models are most compatible with the observed SEDs,
\begin{equation}
\chi^2=\sum_{i=1}^{22}{\frac{[m_{\nu_i}^{\rm intr}-m_{\nu_i}^{\rm
mod}(t)]^2}{\sigma_{i}^{2}}},
\end{equation}
where $m_{\nu_i}^{\rm mod}(t)$ is the integrated magnitude in the
$i^{\rm th}$ filter of a theoretical SSP at age $t$ (for solar
metallicity), $m_{\nu_i}^{\rm intr}$ is the intrinsic, integrated
magnitude, and $\sigma_i$ is the magnitude uncertainty, defined as
\begin{equation}
\sigma_i^{2}=\sigma_{{\rm obs},i}^{2}+\sigma_{{\rm
mod},i}^{2}+(R_{\lambda_i} * \sigma_{\rm red})^{2}+\sigma_{{\rm
md},i}^{2}.
\end{equation}
Here, $\sigma_{{\rm obs},i}$ is the observational uncertainty
from column (6) of Table 1 and column (2) of Table 3, $\sigma_{{\rm
mod},i}$ is the uncertainty associated with the model itself,
$\sigma_{\rm red}$ is the uncertainty in the reddening value, and
$R_{\lambda_i}=A_{\lambda_i}/E(B-V)$, where $A_{\lambda_i}$ is taken
from \citet{car89}, $R_{V}=A_V/E(B-V)=3.1$, and $\sigma_{{\rm md},i}$
is the uncertainty in the distance modulus, for the $i^{\rm th}$
filter. \citet{charlot96} estimated the uncertainty associated with
the term $\sigma_{{\rm mod},i}$ by comparing the colors obtained from
different stellar evolutionary tracks and spectral
libraries. Following \citet{ma07a,ma09}, we adopt
$\sigma_{{\rm mod},i}=0.05$ mag.

\citet{perina09} pointed out that VDB0-B195D is a particularly
extended object and that the photometric measurements of
\citet{Sidney69}, \citet{batt87}, \citet{king91}, and \citet{sh95} did
not include all of its flux. Therefore, we adopt the photometry of
\citet{perina09} to fit the observed SED with theoretical SSPs for our
age determination. The fit yielding the minimum $\chi^2$ value
($\chi^2({\rm min})$) was adopted as the best fit and we adopted the
corresponding age value, $60.0\pm 8.0$~Myr. In addition, our
best-fitting age estimate of $60.0\pm 10.0$ Myr results from using the
(redder) $k$--$p$ and $IJHK_{\rm s}$ photometry; using only the blue
part of the cluster's SED ($B, a$--$e$, where any effects caused by
stochasticity may be smaller) yields an age of $72.0\pm 34.0$ Myr.
The uncertainty was estimated using confidence limits. If
$\chi^2/{\nu}<\chi^2 ({\rm min})/{\nu}+1$, the resulting age is within the
68.3\% probability range; here, ${\nu}=21$ is the number of free
parameters, i.e., the number of observational data points minus the
number of parameters used in the theoretical model. Therefore, the
accepted age range is derived from those fits that have $\chi^2 ({\rm
min})/{\nu}<\chi^2/{\nu}<\chi^2 ({\rm min})/{\nu}+1$. The best
reduced-$\chi^2$---defined as $\chi_{\nu}^2({\rm min})=\chi^2({\rm
min})/{\nu}$---and age are listed in Table 4. The best fit to the SED
of VDB0-B195D is shown in Fig. 4, where we display the intrinsic
cluster SED (symbols with error bars), as well as the integrated SED
(open circles) and spectrum of the best-fitting model. From Fig. 4, we
note that the observational data in the $b$, $d$, $o$, and $p$ BATC
filters and in the $K_{\rm s}$ band do not match the best-fitting
model very well (the difference is approximately 0.3 mag).
Photometric uncertainties in these filters may cause some differences,
although this might not be the main reason for the discrepancy.
As we know, observational star clusters'
SEDs are affected by age, metallicity and reddening. If the reddening
value and metallicity adopted in this paper are not problematic,
discrepancy between our observations and the best-fitting model may
reflect the difficulty in achieving an appropriate (but formal) fit of
an SED of a single, real cluster by SSP models. However, as we will
see below, the reddening value adopted in this paper may be bigger
than the actual reddening of VDB0-B195D. In addition, the
differences between the photometric data and the model in Fig. 4 show
a somewhat systematic behavior with wavelength: in bluer passbands the
cluster seems to be more luminous than predicted by the model, while
in redder passbands it is fainter than the corresponding model
predictions. A blue excess and red deficiency in the observed SED with
respect to the model predictions may indicate a shortage of red giants
(RGs), which can occur when the cluster is either younger or less
massive (or both) than the corresponding best-fitting model suggests.
In other words, IMF discreteness may play a role: due to a relatively
longer main-sequence (MS) phase and shorter RG phase, a random young
cluster is typically bluer than  predicted by SSP models. At the same time,
we find that the reddening value adopted affects the fitting result
greatly. In fact, the best fit to the SED of VDB0-B195D improves a great deal when
adopting a smaller reddening value such as $E(B-V)=0.1$: $\chi_{\nu}^2({\rm min})=0.73$;
the resulting age ($64.0\pm 8.0$ Myr) is nearly the same as one ($60.0\pm 8.0$ Myr)
obtained with $E(B-V)=0.2$.

We next determined the mass of VDB0-B195D. The {\sc galev} models
include absolute magnitudes (in the Vega system) in 77 filters for
SSPs of $10^6~M_\odot$, including 66 filters of the {\sl HST}, Johnson
$UBVRI$ \citep[see for details][]{lan83}, Cousins $RI$ \citep[see for
details][]{lan83}, and $JHK$ \citep{bb88} systems. The difference
between the intrinsic absolute magnitudes and those given by the model
provides a direct measurement of the cluster mass, in units of
$10^6~M_\odot$. However, we should keep in mind that this is only
correct for cluster masses above $10^6~M_\odot$. We estimated the mass of VDB0-B195D
using magnitudes in all of the $BVRI$ and $JHK_{\rm s}$
bands. Therefore, we transformed the 2MASS $JHK_{\rm s}$ magnitudes to
the photometric system of \citet{bb88} using the equations given by
\citet{Carpenter01}. The resulting mass determinations for VDB0-B195D
are listed in Table 5 with their $1\sigma$ uncertainties
including contributions from uncertainties in extinction and distance
modulus. From Table 5, we see that the mass of VDB0-B195D obtained
based on the magnitudes in different filters is very different. (The
highest mass obtained, based on the $B$-band magnitude, is
$~0.5\times10^5~M_\odot$ more massive than that obtained using the
$K_s$ magnitude.) In addition, the mass estimates
differ systematically with filters. Provided that VDB0-B195D is massive enough to be fitted
by SSP models, a systematic trend of
masses based on different passbands may indicate a problem with reddening value adopted for the cluster.
If the actual reddening is smaller than the adopted value, the actual luminosity would be overestimated.
This effect is small in redder filters but strong in bluer filters.
As discussed in age estimation, a smaller reddening value can improve the fitting
result greatly. In fact, a smaller reddening value can reduce the mass discrepancies
based on the magnitudes in different filters. When we adopted $E(B-V)=0.1$, the mass of VDB0-B195D obtained
based on the magnitudes in different filters is the same within 1$\sigma$. We list these estimates
in Table 7. From Table 5, we know that the mass of VDB0-B195D obtained in paper
is between $(1.1-1.6)\times 10^5~M_\odot$ when the reddening value is adopted to be
$(B-V)=0.2$.

\begin{figure*}
\figurenum{4} \epsscale{0.66}\hspace{1.0cm}\rotatebox{-90}{\plotone{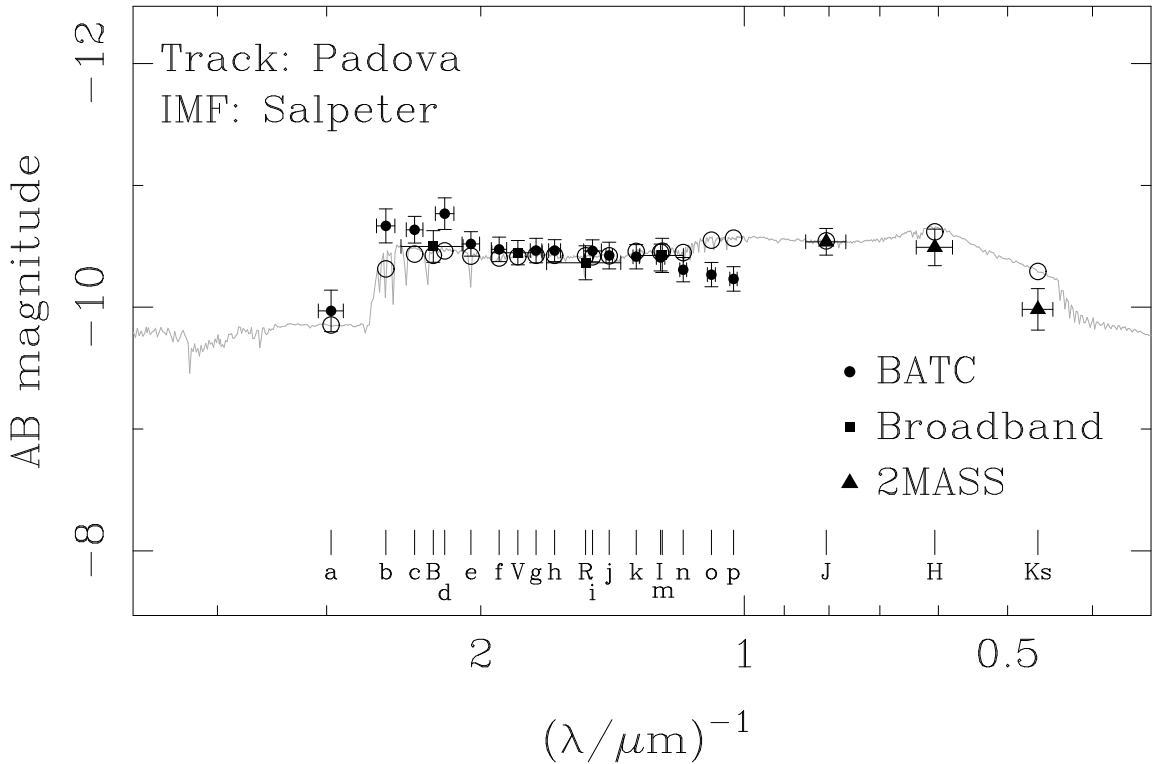}} \vspace{0.5cm}
\caption{Best-fitting, integrated theoretical {\sc galev} SEDs
compared to the intrinsic SED of VDB0-B195D. The photometric
measurements are shown as symbols with error bars (vertical for
uncertainties and horizontal for the approximate wavelength coverage
of each filter). Open circles represent the calculated magnitudes of
the model SED for each filter. We used a distance modulus of
$(m-M)_{0}=24.47$ mag (785 kpc) for M31 \citep{McConnachie05} to
calculate the absolute magnitudes.} \label{fig:four}
\end{figure*}

\section{Summary and discussion}

VDB0-B195D was previously shown to be a massive cluster based on {\sl
HST}/WFPC2 observations. Its color is extremely blue and it is very
bright, particularly in blue bands. In addition, VDB0-B195D is an
extended object, and most previous photometric measurements did not
include its full flux distribution \citep[see][for details]{perina09}.

In this paper, we obtained the cluster's SED in the 15 BATC
intermediate-band filters. We subsequently determined its age and mass
by comparing our multicolor photometry with theoretical stellar
population synthesis models. Our multicolor photometric data
consist of 15 intermediate-band filters obtained in this paper, and
broad-band $BVRI$ and 2MASS $JHK_{\rm s}$ from \citet{perina09},
covering a wavelength range from 3000 to 20,000~\AA. Our results show
that VDB0-B195D is a genuine YMC in M31.

To understand the real nature of the BLCCs, \citet{perina09,perina09b}
performed an {\sl HST} imaging survey of 20 BLCCs in M31's disk. As a
test case, \citet{perina09} presented details of the data-reduction
pipeline that will be applied to all survey data and describe its
application to VDB0-B195D. They estimated the object's age, by
comparison of the observed CMD with theoretical isochrones from
\citet{Girardi02}, at $\simeq 25$ Myr. In addition, they constrained
realistic upper and lower limits to the cluster's age, independent of
the adopted metallicity, within the relatively narrow range from 12 to
63 Myr. Using Maraston's SSP models of solar metallicity
\citep{Maraston98,Maraston05}, \citet{salp55} and \citet{Kroupa01}
IMFs, and photometric values in the $V$ and 2MASS $J$, $H$, and
$K_{\rm s}$ bands, \citet{perina09} concluded that the mass of
VDB0-B195D is $>2.4\times 10^4~M_\odot$, with their best estimates in
the range $\simeq (4-9)\times 10^4~M_\odot$.

\citet{cald09} presented an updated catalog of 1300 objects in M31,
including spectroscopic and imaging surveys, based on images from
either the LGGS or the DSS and spectra taken with the Hectospec fiber
positioner and spectrograph on the 6.5 m MMT. They derived ages and
reddening values for 140 young clusters by comparing their observed
spectra with model spectra from the Starburst99 SSP suite
\citep{Leitherer99}. The results show that these clusters are less
than 2 Gyr old, while most have ages between $10^8$ and $10^9$ yr (the
age of VDB0-B195D they derive is $\log {\rm age/yr} =7.6$). In
addition, \citet{cald09} also estimated the masses of these young
clusters using $V$-band photometry and model mass-to-light ratios
\citep{Leitherer99} corresponding to the derived spectroscopic
ages. This resulted in masses ranging from $2.5 \times 10^2$ to $1.5
\times 10^5~M_\odot$. The mass of VDB0-B195D obtained by
\citet{cald09} is $\log M_{\rm cl}/M_\odot =5.1$ (no uncertainty
quoted).

We compare the various age and mass estimates of VDB0-B195D in Table
6. Our newly obtained age is older than the estimates of both
\citet{perina09} and \citet{cald09}, while the mass obtained in this
paper is higher than the estimate of \citet{perina09} and consistent
with the determination of \citet{cald09}. However, our results are in
agreement with those of both \citet{perina09} and \citet{cald09}
within $3\sigma$. The age and mass obtained in this paper confirms
that VDB0-B195D is genuinely a YMC in M31.

As we know, SSP models describe a very special case of a
continuous distribution of stellar mass (or light) along isochrones. This is well
approximated by clusters with masses larger than $10^6~M_\odot$. Also, for cluster
masses of about $10^5~M_\odot$,
SSP models can probably still be applied since a systematic difference between
SSP models and observations should, on average, be smaller than 0.05 mag for
clusters older than 10 Myr (see Fig. 3 in Piskunov et al. 2009).
However, from the results of this paper, we may conclude that, probably, a formal fitting of SSP
models to observed SEDs cannot be used without caution even for relatively
massive (or apparently massive) clusters, and it is highly doubtful that
this approach can be applied in a routine work providing accurate cluster
parameters. The relative accuracy of 10\% for age and 20\% found for the mass of
VDB0-B195D seems to be rather formal and not very confident. In addition,
observational star clusters'
SEDs are affected by reddening, an effect that is also difficult to
separate from the combined effects of age and metallicity (Calzetti
1997; Vazdekis et al. 1997; Origlia et al. 1999). Only the
metallicity and reddening are derived accurately (and, ideally, independently),
these degeneracies are largely (if not entirely) reduced,
and ages can then also be estimated accurately based on a comparison
of multicolor photometry spanning a significant wavelength
range (de Grijs et al. 2003b; Anders et al. 2004) with theoretical
stellar population synthesis models. It is true that the
discrepancy between our observations and the best-fitting model is great, and
the mass of VDB0-B195D obtained
based on the magnitudes in different filters is very different.
However, when we adopt a smaller reddening value, the results improve
greatly. So, we conclude that the actual reddening value of VDB0-B195D may
be smaller than $E(B-V)=0.2$.

\acknowledgments We are indebted to the anonymous referee for very carefully
reading our manuscript, and for many thoughtful comments and insightful
suggestions that improved this paper significantly. We are grateful to Dr. Richard de Grijs for
the help in terms of scientific input and proofreading. This work was supported by the Chinese
National Natural Science Foundation (grants 10873016, 10633020, 10603006, and 10803007) and by the
National Basic Research Program of China (973 Program), No. 2007CB815403.

\newpage
\begin{table}
\begin{center}
\caption{BATC photometry of the M31 YMC VDB0-B195D.} \label{t1.tab}
\begin{tabular}{cccccc}
\tableline \tableline
Filter & Central wavelength & Bandwidth & Number of images & Exposure time  &Magnitude \\
       &  (\AA)             &  (\AA)    &                  & (hours)        &          \\
\tableline
$a$& 3360 & 222 & 6&  2.0 & $15.51\pm 0.14$\\
$b$& 3890 & 187 & 6&  2.0 & $14.73\pm 0.11$\\
$c$& 4210 & 185 & 3&  0.8 & $14.70\pm 0.07$\\
$d$& 4550 & 222 & 3&  1.0 & $14.49\pm 0.10$\\
$e$& 4920 & 225 & 3&  1.0 & $14.67\pm 0.05$\\
$f$& 5270 & 211 & 3&  1.0 & $14.65\pm 0.05$\\
$g$& 5795 & 176 & 3&  1.0 & $14.59\pm 0.04$\\
$h$& 6075 & 190 & 3&  1.0 & $14.56\pm 0.02$\\
$i$& 6660 & 312 & 3&  1.0 & $14.50\pm 0.02$\\
$j$& 7050 & 121 & 5&  1.7 & $14.51\pm 0.06$\\
$k$& 7490 & 125 & 3&  1.0 & $14.47\pm 0.05$\\
$m$& 8020 & 179 & 3&  1.0 & $14.43\pm 0.07$\\
$n$& 8480 & 152 & 6&  2.0 & $14.49\pm 0.05$\\
$o$& 9190 & 194 & 6&  2.0 & $14.49\pm 0.05$\\
$p$& 9745 & 188 & 6&  2.0 & $14.50\pm 0.05$\\
\tableline
\end{tabular}\\
\end{center}
\end{table}

\begin{table}
\begin{center}
\caption{Comparison of broad-band photometry of VDB0-B195D.}
\label{t2.tab}
\begin{tabular}{lcccccc}
\tableline \tableline
Filter & ${\rm Mag}^{a}$ & ${\rm Mag}^{b}$ & ${\rm Mag}^{c}$ & ${\rm Mag}^{d}$ & ${\rm Mag}^{e}$ & ${\rm Mag}^{f}$\\
\tableline
   $U$   & 14.66  & 15.11 & $15.12\pm 0.012$ & $14.97\pm 0.01$ & 15.110 & 15.140 \\
   $B$   & 15.14  & 15.39 & $15.49\pm 0.010$ & $15.31\pm 0.01$ & 15.410 & 15.510 \\
   $V$   & 14.94  & 15.19 & $15.32\pm 0.013$ & $15.06\pm 0.01$ & 15.190 & 15.280 \\
   $R$   &        & 15.27 &                  &                 & 14.920 &        \\
\tableline
\end{tabular}\\
\end{center}
{\hspace{4.2cm}$^{a}${\citet{Sidney69}};
$^{b}${\citet{batt87}};
$^{c}${\citet{king91}, uncertainties are the median uncertainties in the mean for all sample cluster measurements;}
$^{d}${\citet{sh95}};
$^{e}${Photometry from \citet{gall04}, based on \citet{batt87};}
$^{f}${Photometry from \citet{gall04}, based on \citet{sh95}.}}
\end{table}

\begin{table}
\begin{center}
\caption{Recently determined photometry for VDB0-B195D.} \label{t3.tab}
\begin{tabular}{lccc}
\tableline \tableline
Filter & ${\rm Mag}^{a}$ & ${\rm Mag}^{b}$ & ${\rm Mag}^{c}$ \\
\tableline
   $U$   &                  &                 & $14.37\pm 0.22$ \\
   $B$   &  $14.94\pm 0.09$ &                 & $14.62\pm 0.13$ \\
   $V$   &  $14.67\pm 0.05$ &                 & $14.67\pm 0.05$ \\
   $R$   &  $14.45\pm 0.11$ &                 & $14.60\pm 0.06$ \\
   $I$   &  $14.01\pm 0.11$ &                 & $14.19\pm 0.10$ \\
   $J$   &  $13.26\pm 0.07$ & $13.78\pm 0.03$ &   \\
   $H$   &  $12.76\pm 0.12$ & $13.15\pm 0.04$ &   \\
   $K_s$ &  $12.77\pm 0.15$ & $12.96\pm 0.03$ &   \\
\tableline
\end{tabular}\\
\end{center}
{\hspace{5.5cm}$^{a}${\citet{perina09}};
$^{b}${\citet{gall04}};
$^{c}${This paper}.}
\end{table}

\begin{table}
\begin{center}
\caption{Age estimate of VDB0-B195D based on the the {\sc
galev} models.} \label{t4.tab}
\begin{tabular}{ccc}
\tableline \tableline
  Age          &  log (Age)     &   $\chi_{\nu}^2({\rm min})$     \\
  (Myr)        &  [yr]         &   (per degree of freedom)\\
\hline
$60.0\pm8.0$ & $ 7.78\pm0.05$ &    2.2                 \\
\tableline
\end{tabular}
\end{center}
\end{table}

\begin{table}
\begin{center}
\caption{Mass estimates (and uncertainties) of
  VDB0-B195D based on the {\sc galev} models.} \label{t5.tab}
%\begin{tabular}{ccccccc|ccc}
\begin{tabular}{ccccccc}
\tableline \tableline
%$B$ & $V$ & $R$ & $I$                  & $J$ & $H$ & $K_{\rm s}$ &                    & Mean value &     \\
$B$ & $V$ & $R$ & $I$                  & $J$ & $H$ & $K_{\rm s}$ \\
\tableline
%    &     &     & Mass $(10^5~M_\odot)$ &     &     &    & Mass $(10^5~M_\odot)$&    & log (Mass/$M_\odot$)\\
    &     &     & Mass $(10^5~M_\odot)$ &     &     &    \\
\tableline
%1.64  & 1.62  & 1.42  & 1.37  & 1.41  &  1.28 & 1.09  & {\bf 1.37}  &  & {\bf 5.14} \\
%0.18  & 0.13  & 0.17  & 0.16  & 0.13  & 0.16  & 0.17  & {\bf 0.03}  &  & {\bf 0.01} \\
$1.6\pm0.18$  & $1.6\pm0.13$  & $1.4\pm0.17$  & $1.4\pm0.16$  & $1.4\pm0.13$ &  $1.3\pm0.16$ & $1.1\pm0.17$  \\
\tableline
\end{tabular}
\end{center}
\end{table}

\begin{table}
\begin{center}
\caption{Comparison of age and mass estimates of VDB0-B195D.}
\label{t6.tab}
\begin{tabular}{cccc|cccc}
\tableline \tableline
  Age$^1$ &  Age$^2$ & log (Age)$^3$ & log (Age)$^2$  &  Mass$^1$         &   Mass$^2$        & log (Mass)$^3$ & log (Mass)$^2$     \\
  (Myr)         &     (Myr)      &   [yr]             &   [yr]              &  ($10^4~M_\odot$) & ($10^5~M_\odot$)  &[$M_\odot$]     & [$M_\odot$]\\
\hline
%25           & $60.0\pm8.0$ & 7.6                &$ 7.78\pm0.05$        &  $4-9$                  & $1.37\pm0.03$& $5.1$     &   $5.14\pm0.01$ \\
25           & $60.0\pm8.0$ & 7.6                &$ 7.78\pm0.05$        &  $4-9$                  & $1.1-1.6$ & $5.1$     &   $5.0-5.2$ \\
\tableline
\end{tabular}\\
\end{center}
{\hspace{2.5cm}$^{1}${\citet{perina09}}; $^{2}${This paper}; $^{3}${\citet{cald09}.}\\}
\end{table}

\begin{table}
\begin{center}
\caption{Mass estimates (and uncertainties) of
  VDB0-B195D based on the {\sc galev} models with $E(B-V)=0.1$.} \label{t7.tab}
%\begin{tabular}{ccccccc|ccc}
\begin{tabular}{ccccccc}
\tableline \tableline
%$B$ & $V$ & $R$ & $I$                  & $J$ & $H$ & $K_{\rm s}$ &                    & Mean value &     \\
$B$ & $V$ & $R$ & $I$                  & $J$ & $H$ & $K_{\rm s}$ \\
\tableline
%    &     &     & Mass $(10^5~M_\odot)$ &     &     &    & Mass $(10^5~M_\odot)$&    & log (Mass/$M_\odot$)\\
    &     &     & Mass $(10^5~M_\odot)$ &     &     &    \\
\tableline
$1.1\pm0.12$  & $1.2\pm0.10$  & $1.1\pm0.14$  & $1.2\pm0.14$  & $1.3\pm0.12$  & $1.2\pm0.16$  & $1.1\pm0.17$  \\
\tableline
\end{tabular}
\end{center}
\end{table}

\end{document}